\documentclass[sigconf,authorversion,]{acmart} 

\usepackage{url}
\newcommand{\blueurl}[1]{\urlstyle{same}\color{blue}\url{#1}\color{black}\urlstyle{tt}}
\AtBeginDocument{%
  \providecommand\BibTeX{{%
    \normalfont B\kern-0.5em{\scshape i\kern-0.25em b}\kern-0.8em\TeX}}}

\setcopyright{none}

\acmConference[Preprint]{2024}{March}

\begin{document}

\title{ACM MMSys 2024 Bandwidth Estimation in Real Time Communications Challenge}

\author{Sami Khairy}
\affiliation{%
  \institution{Microsoft Corporation}
  \city{Vancouver}
  \country{Canada}
}
\email{samikhairy@microsoft.com}

\author{Gabriel Mittag}
\affiliation{%
  \institution{Microsoft Corporation}
  \city{Redmond}
  \country{USA}
}
\email{gmittag@microsoft.com}

\author{Vishak Gopal}
\affiliation{%
  \institution{Microsoft Corporation}
  \city{Redmond}
  \country{USA}
}
\email{vishak.gopal@microsoft.com}

\author{Francis Y. Yan}
\affiliation{%
  \institution{Microsoft Research}
  \city{Redmond}
  \country{USA}
}
\email{francisy@microsoft.com}

\author{Zhixiong Niu}
\affiliation{%
  \institution{Microsoft Research}
  \city{Beijing}
  \country{China}
}
\email{zhniu@microsoft.com}

\author{Ezra Ameri}
\affiliation{%
  \institution{Microsoft Corporation}
  \city{Vancouver}
  \country{Canada}
}
\email{ezraameri@microsoft.com}

\author{Scott Inglis}
\affiliation{%
  \institution{Microsoft Corporation}
  \city{Redmond}
  \country{USA}
}
\email{singlis@microsoft.com}

\author{Mehrsa Golestaneh}
\affiliation{%
  \institution{Microsoft Corporation}
  \city{Toronto}
  \country{Canada}
}
\email{mgolestaneh@microsoft.com}

\author{Ross Cutler}
\affiliation{%
  \institution{Microsoft Corporation}
  \city{Redmond}
  \country{USA}
}
\email{ross.cutler@microsoft.com}

\renewcommand{\shortauthors}{Khairy, et al.}

\begin{abstract}
  The quality of experience (QoE) delivered by video conferencing systems to end users depends in part on correctly estimating the capacity of the bottleneck link between the sender and the receiver over time. Bandwidth estimation for real-time communications (RTC) remains 
  a significant challenge, primarily due to the continuously evolving heterogeneous network architectures and technologies. From the first bandwidth estimation challenge which was hosted at ACM MMSys 2021, we learned that bandwidth estimation models trained with reinforcement learning (RL) in simulations to maximize network-based reward functions may not be optimal in reality due to the sim-to-real gap and the difficulty of aligning network-based rewards with user-perceived QoE. This grand challenge aims to advance bandwidth estimation model design by aligning reward maximization with user-perceived QoE optimization using offline RL and a real-world dataset with objective rewards which have high correlations with subjective audio/video quality in Microsoft Teams. All models submitted to the grand challenge underwent initial evaluation on our emulation platform. For a comprehensive evaluation under diverse network conditions with  temporal fluctuations, top models were further evaluated on our geographically distributed testbed by using each model to conduct $600$ calls within a $12$-day period. The winning model is shown to deliver comparable performance to the top behavior policy in the released dataset. By leveraging real-world data and integrating objective audio/video quality scores as rewards, offline RL can therefore facilitate the development of competitive bandwidth estimators for RTC.
\end{abstract}



\maketitle

\section{Introduction}

Video conferencing systems have recently emerged as indispensable tools to sustain global business operations and enable accessible education by revolutionizing the way people connect, collaborate, and communicate despite physical barriers and geographical divides \cite{markudova2023recoco,eo2022opennetlab}. The quality of experience (QoE) delivered by these systems to the end user depends in part on bandwidth estimation, which is the problem of estimating the varying capacity of the bottleneck link between the sender and the receiver over time \cite{bentaleb2022bob}. In real time communication systems (RTC), the bandwidth estimate serves as a target bit rate for the audio/video encoder, controlling the send rate from the client \cite{li2022reinforcement,wang2021hybrid}. Overestimating the capacity of the bottleneck link causes network congestion as the client sends data at a rate higher than what the network can handle \cite{zhang2020onrl}. Network congestion is characterized by increased delays in packet delivery, jitter, and potential packet losses. In terms of user’s experience, users will typically experience many resolution switches, frequent video freezes, garbled speech, and audio/video desynchronization, to name a few. Underestimating the available bandwidth on the other hand causes the client to encode and transmit the audio/video streams in a lower rate signal than what the network can handle, which leads to underutilization and degraded QoE. Estimating the available bandwidth accurately is therefore critical to providing the best possible QoE to users in RTC systems. Nonetheless, accurate bandwidth estimation is faced with a multitude of challenges due to dynamic network paths between senders and receivers with fluctuating traffic loads, existence of diverse wired and wireless heterogeneous network technologies with distinct characteristics, presence of different transmission protocols fighting for bandwidth to carry side and cross traffic, and partial observability of the network as only local packet statistics are available at the client side to base the estimate on.  

In the previous bandwidth estimation challenge that was hosted at ACM MMSys 2021 (\blueurl{https://2021.acmmmsys.org/rtc_challenge.php}), participants were provided with a “gym” simulation environment based on network simulator 3 (NS-3) and the challenge focused on learning a bandwidth estimator using online reinforcement learning (RL). Policies trained in simulation with network-based reward functions may not be optimal when deployed in the real world because of many challenges, including, sim-to-real gap \cite{9828569}, and the misalignment between rewards computed based on network measurements and actual user perceived quality of experience \cite{clcc}. To improve QoE for users in RTC systems, the ACM MMSys 2024 grand challenge aims to advance the field of bandwidth estimation for RTC by proposing to train bandwidth estimation models through offline RL using a real-world dataset consisting of observed network dynamics and objective metrics which are highly correlated with user-perceived audio/video quality in Microsoft Teams.

\section{Challenge Description}

Offline RL tackles the problem of learning effective control policies from a static dataset of previously collected experiences \cite{fujimoto2019off,lange2012batch,singh2024reds,fu2022closer}, eliminating the need for online environment interaction \cite{sutton2018reinforcement}. The offline RL framework is particularly useful in many real-world control problems, such as bandwidth estimation, where generating data through online interaction is not only expensive and time-consuming, but also potentially dangerous and impractical due to the risks associated with executing untrained exploratory policies in a real control system. While favourable for many reasons, offline RL is faced with the challenge of trading-off between two conflicting objectives: learning a policy whose performance improves upon that of the behaviour policy that is used to collect the dataset, and minimizing deviations from the behaviour policy to avoid out of distribution actions that can be catastrophic. Recent advancements in offline RL algorithm design tackle this challenge in different ways, such as constraining the policy implicitly or explicitly \cite{fujimoto2021minimalist,fujimoto2019off,kumar2019stabilizing,wu2019behavior}, or regularizing the Q or value function to predict low values for out of distribution actions \cite{kostrikov2021offline, kumar2020conservative,nakamoto2024cal,wang2020critic}. We believe that the advancements in offline RL make it a viable technology to advance the field of bandwidth estimation in RTC \cite{gottipati2023real, yen2023computers}.

In this challenge, participants are provided with a dataset of trajectories based on real-world Microsoft Teams audio/video calls. Each trajectory corresponds to an audio/video call between a pair of machines. During a call, each machine transmits an audio and a video stream to the other machine. The dataset includes objective signals for audio and video quality, which quantify the perceived quality of the received audio and video streams by the user. These signals are predicted by ML models whose predictions have high correlation with subjective audio and video quality scores as determined by ITU-T’s $P.808$ and $P.910$, respectively \cite{mittag2023lstm}.

The goal of the challenge is to improve QoE for RTC system users as measured by objective audio/video quality scores by developing a deep learning-based policy model (receiver-side bandwidth estimator, $\pi$) with offline RL. To this end, participants are free to specify an appropriate reward function based on the provided dataset of observed network dynamics and objective metrics, the model architecture, and the training algorithm, given that the developed model adheres to the challenge requirements detailed on the challenge website (\blueurl{https://www.microsoft.com/en-us/research/academic-program/bandwidth-estimation-challenge/}).

\section{Challenge Datasets}

Our GitHub repository\footnote{\blueurl{https://github.com/microsoft/RL4BandwidthEstimationChallenge}} open sourced two datasets for Microsoft Teams audio/video calls. The first dataset\footnote{\blueurl{https://github.com/microsoft/RL4BandwidthEstimationChallenge/blob/main/download-testbed-dataset.sh}} is generated from $18859$ audio/video peer-to-peer (P2P) Microsoft Teams calls conducted between testbed nodes which are geographically distributed across many countries and continents. Testbed nodes are connected to the internet through various Internet Service Providers (ISPs) over either wired or wireless connections. Because the performance of policies trained with offline RL depends heavily on the coverage of behaviour in the dataset \cite{kumar2022should}, calls have been conducted with six bandwidth estimators (behaviour policies), including traditional methods such as Kalman-filtering-based estimators and WebRTC (Web Real Time Communications), as well as different ML policies. The behaviour policies are code-named $\{v_0,v_1 \cdots, v_5\}$ in the dataset. 

On the other hand, the second dataset\footnote{\blueurl{https://github.com/microsoft/RL4BandwidthEstimationChallenge/blob/main/download-emulated-dataset.sh}} is generated 
from $9405$ test calls conducted between pairs of machines that are connected through a networking emulation software which emulates different network characteristics such as burst loss, traffic policing, and bandwidth fluctuations, to name a few. The characteristics of the bottleneck link, namely ground truth capacity and loss rate, are randomly varied throughout the duration of the test call to generate a diverse set of trajectories with network dynamics that may not occur in the real world but are nevertheless important to enhance state-action space coverage and aid in learning generalizable policies \cite{kumar2022should}. Because this dataset is generated through emulation, it contains ground truth information about the bottleneck link capacity and loss rate.

In either of these two datasets, each audio/video call leg is represented as a trajectory consisting of a sequence of quadruples $(o_n, a_n, r_n^\text{audio}, r_n^\text{video})$, where $o_n$ is a high-dimensional observation vector computed based on packet information received by the client, $a_n$ is the predicted bandwidth in bits-per-second (bps), and $r_n^\text{audio},~r_n^\text{video}$ are the audio and video quality reward signals, respectively. $r_n^\text{audio},~r_n^\text{video}$ are predicted by reference-free and reference-based deep learning (DL) models which map audio/video streams to the mean opinion score (MOS) $\in [0,5]$, with a score of $5$ being the highest. These models attain high ($>0.95$) Pearson correlation coefficient (PCC) with subjective audio and video MOS
from crowdsourcing audio/video-quality experiments.

The observation vector $o_n \in \mathbb{R}^{150}$ at a time step $n$ is encapsulates observed network statistics that characterize the state of the bottleneck link between the sender and receiver over the $5$ most recent short term monitor intervals (MI) of $60$ms and the $5$ most recent long-term MIs of $600$ms. Specifically, the observation vector tracks $15$ different network features computed based on RTP \cite{schulzrinne2003rtp} packet header information received by the client over $5$ short and $5$ long term MIs ($15$ features $\times$ ($5$ short term MIs + $5$ long term MIs) = $150$) as follows:

\begin{enumerate}
    \item Receiving rate: rate at which the client receives data from the sender during a MI, unit: bps.
    \item Number of received packets: total number of packets received in a MI, unit: packet.
    \item Received bytes: total number of bytes received in a MI, unit: Bytes.
    \item Queuing delay: average delay of packets received in a MI minus the minimum packet delay observed so far, unit: ms.
    \item Delay: average delay of packets received in a MI minus a fixed base delay of 200ms, unit: ms.
    \item Minimum seen delay: minimum packet delay observed so far, unit: ms.
    \item Delay ratio: average delay of packets received in a MI divided by the minimum delay of packets received in the same MI, unit: ms/ms.
    \item Delay average minimum difference: average delay of packets received in a MI minus the minimum delay of packets received in the same MI, unit: ms.
    \item Packet interarrival time: mean interarrival time of packets received in a MI, unit: ms.
    \item Packet jitter: standard deviation of interarrival time of packets received in a MI, unit: ms.
    \item Packet loss ratio: probability of packet loss in a MI, unit: packet/packet.
    \item Average number of lost packets: average number of lost packets given a loss occurs, unit: packet.
    \item Video packets probability: proportion of video packets in the packets received in a MI, unit: packet/packet.
    \item Audio packets probability: proportion of audio packets in the packets received in a MI, unit: packet/packet.
    \item Probing packets probability: proportion of probing packets in the packets received in a MI, unit: packet/packet.
\end{enumerate}

The 5 short term MI features are indexed at $\{$(feature $\#$ – 1) $\times$ 10, $\cdots$, feature $\#$ $\times$ 10 – 5 – 1$\}$. On the other hand, the 5 long term MI features are indexed at $\{$feature $\#$ $\times$ 10 – 5, $\cdots$, feature $\#$ $\times$ 10 – 1$\}$.

Participants are free to stratify or split the open-sourced datasets into train/validation sets as deemed necessary for training a bandwidth estimator policy. However, the ground truth information in the emulated dataset cannot be used as inputs to the model since this information is not available outside the lab setup and the testing environment is only partially observable. Ground truth information can be used for exploratory data analysis, model selection, as part of an auxiliary prediction task, or in an asymmetric actor-critic RL setup, where the critic can have extra information during training.

\section{Evaluation Setup}

Along with the training datasets, we have open-sourced a lightweight baseline model\footnote{\blueurl{https://github.com/microsoft/RL4BandwidthEstimationChallenge/tree/main/onnx_models}} trained only on the emulated dataset with Implicit Q-Learning (IQL) \cite{kostrikov2021offline}. The model consists of a normalization layer to standardize the input features, followed by $2$ dense layers with 128 neurons each and $\text{tanh}$ activation function. The final layer predicts the mean and standard deviation of the action $\hat{a}_n \in [-1,1]$. The predicted action mean is finally transformed to bps using the transformation,

\begin{equation}
    a_n = \text{exp}\Big( \frac{\hat{a}_n+1}{2} \times \text{ln}(800) + \text{ln}(0.01)\Big)\times 10^6.
\end{equation}

The baseline model was trained with a discount factor of $0.99$, learning rate of $3e-4$, batch size of $16384$ samples, temperature parameter of $8$, and an expectile regression parameter of $0.7$.

\subsection{Preliminary Evaluation Opportunities}

To support the research efforts of challenge participants and enhance the overall quality of submissions ahead of the final deadline, we have offered \textbf{two optional} preliminary evaluation opportunities for all registered teams. Participants had the chance to submit up to three models in the first preliminary evaluation opportunity and up to two models in the second preliminary evaluation opportunity for online testing.  This initiative aimed to assist participants with refining their designs, identifying potential flaws early in the process, and ultimately enhancing the robustness of the solutions.

Each model submitted to either of these preliminary evaluation opportunities was evaluated in our emulation platform by conducting $24$ P2P test calls with $8$ different network traces. The average objective audio/video quality scores for these models were shared with the participants and were posted in a leaderboard on the challenge website\footnote{\blueurl{https://www.microsoft.com/en-us/research/academic-program/bandwidth-estimation-challenge/results/}}.

\subsection{Final Evaluation Methodology}

Final models submitted to the grand challenge were evaluated in a 2-stage evaluation process. In the first stage, all submitted models were evaluated on our emulation platform by conducting $160$ $2$-minute P2P test calls with $16$ different network scenarios. The $16$ network scenarios spanned fixed low bandwidth traces, fixed high bandwidth traces, fluctuating bandwidth traces, burst loss traces, and fluctuating burst loss traces. The purpose of the first evaluation stage on our emulation platform was to obtain an initial ranking and determine the top models which would advance to the second and final evaluation stage.

In the final evaluation stage, the top $3$ models from the first stage were evaluated on our intercontinental testbed by conducting $600$ $3$-minute calls during a $12$-day period from February 14th to February 26th, 2024. These calls were between random pairs of nodes which are geographically distributed across the globe. This comprehensive evaluation stage represents a real-world test of top bandwidth estimators across diverse network conditions with temporal fluctuations over the internet. As per the rules of the grand challenge, the winner and runner-up are determined based on the rankings in the final evaluation stage.

In both evaluation stages, the scoring function $\mathbb{S}$ which has been used to rank the models is, 
\begin{equation}
    \mathbb{S} = \mathbb{E}_\text{call legs}\Bigg[\mathbb{E}_n\Big[{r_n^{\text{audio}} + r_n^{\text{video}}}\Big]\Bigg] \in [0,10],
\end{equation}
where the inner expectation is the temporal average of the objective audio/video reward in a call leg.

\section{Evaluation Results \& Discussion}

\begin{table*}
  \caption{Model prediction accuracy $(mse, e^+, e^-)$ by network scenario (top model in green, second top model in yellow).}
  \label{tab:Scenarioacc}
  \resizebox{\textwidth}{!}{
  \begin{tabular}{|l|c|c|c|c|c|c|}
    \hline
    \textbf{Model} & \textbf{Low BW} & \textbf{High BW} & \textbf{Fluctuating BW} & \textbf{Burst Loss} & \textbf{Fluctuating BL}\\
    \hline
    \hline
Baseline	&  \colorbox{green}{0.0024}, 0.0181, \colorbox{yellow}{0.0881} &  \colorbox{yellow}{4.7594}, \colorbox{yellow}{0.0112}, \colorbox{yellow}{0.2395} &  0.2931, 0.0818, 0.3036 &  \colorbox{yellow}{2.3518}, \colorbox{green}{0.0012}, 0.1993 &  \colorbox{yellow}{3.6386}, \colorbox{green}{0.0007}, \colorbox{yellow}{0.2131} \\  \hline
CUC Echoes	&  0.0153, 0.0219, 0.3800 &  7.3816, 0.0119, 0.3330 &  0.4147, 0.0429, 0.4068 &  2.5595, 0.0066, 0.3568 &  5.2707, 0.0235, 0.2218 \\  \hline
Fast and furious &  0.0111, 0.1290, 0.0984 &  5.7777, 0.0222, 0.2783 &  \colorbox{green}{0.1141}, 0.1950, \colorbox{green}{0.1467} &  3.1843, 0.0631, \colorbox{yellow}{0.1786} &  5.4722, 0.0486, 0.2329 \\  \hline
MediaLab	&  0.0494, 0.0370, 0.7277 &  9.0791, 0.1274, 0.3613 &  0.5409, 0.0783, 0.5441 &  13.1523, 0.0378, 0.7859 &  4.0101, 0.0403, 0.3475 \\  \hline
Paramecium &  0.0043, 0.0151, 0.1472 &  7.7142, 0.0218, 0.3044 &  \colorbox{yellow}{0.1230}, 0.0712, \colorbox{yellow}{0.1822} &  4.8861, 0.0254, 0.2371 &  7.1097, 0.0150, 0.2835 	\\  \hline
SJTU Medialab	&  0.0112, \colorbox{green}{0.0005}, 0.3069 &  7.1261, 0.0252, 0.3112 &  0.4024, \colorbox{green}{0.0262}, 0.4111 &  2.9328, \colorbox{yellow}{0.0042}, 0.2796 &  4.6373, 0.0021, 0.2797 	\\  \hline
Schaferct	&  \colorbox{yellow}{0.0036}, 0.0933, \colorbox{green}{0.0848} &  \colorbox{green}{2.4350}, 0.0338, \colorbox{green}{0.1288} &  0.2543, 0.1393, 0.2261 &  \colorbox{green}{1.4141}, 0.0476, \colorbox{green}{0.1118} &  \colorbox{green}{2.4750}, 0.0275, \colorbox{green}{0.1558} 	\\  \hline
TEN TMS &  0.0135, \colorbox{yellow}{0.0031}, 0.3425 &  12.5444, \colorbox{green}{0.0061}, 0.4406 &  0.4422, \colorbox{yellow}{0.0341}, 0.4011 &  7.9384, 0.0068, 0.3771 &  11.4860, \colorbox{yellow}{0.0014}, 0.3354 	\\
  \hline
\end{tabular}}
\end{table*}

\begin{table*}
  \caption{Model score by network scenario (top model in green, second top model in yellow).}
  \label{tab:ScenarioScore}
  \begin{tabular}{|l|c|c|c|c|c|c|}
    \hline
    \textbf{Model} & \textbf{Low BW} & \textbf{High BW} & \textbf{Fluctuating BW} & \textbf{Burst Loss} & \textbf{Fluctuating BL}\\
    \hline
    \hline
Baseline	&	 \colorbox{green}{6.99±0.08}	&	8.78±0.17	&	6.77±0.23	&	\colorbox{yellow}{7.65±0.17}	&	\colorbox{yellow}{7.92±0.10}	\\  \hline
CUC Echoes	&	6.41±0.29	&	8.47±0.63	&	6.58±0.26	&	7.22±0.39	&	7.71±0.27	\\  \hline
Fast and furious	&	6.54±0.14	&	8.84±0.10	&	6.70±0.26	&	7.47±0.13	&	7.72±0.10	\\  \hline
MediaLab	&	5.40±0.29	&	7.98±1.53	&	5.97±0.59	&	5.85±0.84	&	7.35±0.82	\\  \hline
Paramecium	&	\colorbox{yellow}{6.82±0.05}	&	\colorbox{yellow}{8.86±0.06}	&	\colorbox{green}{6.92±0.26}	&	7.17±0.26	&	7.89±0.11	\\  \hline
SJTU Medialab	&	6.71±0.21	&	8.56±0.41	&	6.78±0.24	&	7.46±0.37	&	7.83±0.21	\\  \hline
Schaferct	&	6.47±0.13	&	\colorbox{green}{8.90±0.05}	&	6.61±0.22	&	\colorbox{green}{7.73±0.08}	&	\colorbox{green}{7.97±0.07}	\\  \hline
TEN TMS	&	6.53±0.24	&	8.23±0.75	&	\colorbox{yellow}{6.88±0.22}	&	7.23±0.67	&	7.53±0.61	\\
  \hline
\end{tabular}
\end{table*}

Out of $21$ teams who have registered for the grand challenge, we have received final submissions from $7$ participating teams. 
Below, we provide a concise overview of each submitted model, highlighting the key features for its training.

\begin{enumerate}
    \item \textbf{Fast and furious} \cite{fastandfurious}: an actor-critic model which is trained in 2-stages. In the first stage, the critic is independently trained to predict audio and video quality scores. In the second stage, the actor is trained on the emulated dataset in an asymmetric approach which leverages the pre-trained critic and ground truth information in the emulated dataset. Objective audio/video MOS scores are used as rewards.
    
    \item \textbf{Schaferct} \cite{schaferct}: an actor-critic model which has a GRU cell and residual connections in its architecture. The model has been trained with IQL on about $10\%$ of the testbed dataset. Objective audio/video MOS scores are used as rewards.

    \item \textbf{TEN-TMS}: a model which consists of an encoder-decoder block that is pre-trained to extract network-aware feature representations, and a multi-expert bandwidth estimation 
    block that is trained with IQL. An accuracy-based data sampling strategy and a curriculum learning approach have been adopted to train the model. An accuracy-based metric is used as a reward.

    \item \textbf{SJTU MediaLab} \cite{pioneer}: an actor-critic model with a variational auto-encoder (VAE) to learn latent representations of network observation vectors and bandwidth estimates. Model design and training are based on BCQ \cite{fujimoto2019off} and TD3+BC \cite{fujimoto2021minimalist}. A network-metric-based function is used as a reward.

    \item \textbf{CUC Echoes}: an actor-critic model which is trained in three stages. First, behaviour cloning (BC) is used to learn a policy based on the dataset. Second, a Q-function is trained to predict the objective audio/video quality scores based on the dataset and using the learned BC policy. Third, the BC policy is improved with a one-step constrained policy improvement step using the trained Q-function. Objective audio/video MOS scores are used as reward.

    \item \textbf{Paramecium} \cite{paramecium}: a rule-based bandwidth estimation model that combines delay-based and rate-based congestion control strategies to maintain a small bottleneck queuing delay. No reward function is used for the design.
     \item \textbf{MediaLab}: an actor-critic model which has been trained with CQL.
\end{enumerate}

\subsection{First Evaluation Stage Results}

The first evaluation stage is conducted on our emulation platform where ground truth information is available. Hence, we focus on analyzing the performance of the models in terms of three metrics that assess prediction accuracy in order to understand how prediction accuracy impacts objective audio and video quality scores. The three metrics are, 1) mean squared error (MSE),
\begin{equation}
    mse = \mathbb{E}\Big[ (a_n - c_n)^2 \Big],
\end{equation}
where $a_n$ and $c_n$ are the predicted bandwidth and ground truth capacity, respectively, 2) overestimation error rate, 
\begin{equation}
    e^+ = \mathbb{E}\Big[ \text{max}\big(0, \frac{a_n-c_n}{c_n}\big)\Big],
\end{equation}
and underestimation error rate,
\begin{equation}
    e^- = \mathbb{E}\Big[ \text{max}\big(0, \frac{c_n-a_n}{c_n}\big)\Big].
\end{equation}
Generally, lower values for these metrics are better. We present a breakdown of these metrics, in addition to the average audio/video quality score achieved by each model in different network scenarios in tables \ref{tab:Scenarioacc} and \ref{tab:ScenarioScore}, respectively. The top and second top performing models in each group of network traces are highlighted in green and yellow, respectively. Based on these results, we make the following two observations, 

\begin{enumerate}
    \item While accurately predicting the ground truth capacity of the bottleneck link can sometimes lead to  better audio/video quality scores, it does not guarantee optimal QoE. For example, Schaferct ranks second ($\#2$) and first ($\#1$) in mse and underestimation error rate ($e^-$) in the low bandwidth scenario, respectively, yet it fails to be among the top two positions in this group. A similar observation holds for the Fast and furious model in fluctuating bandwidth scenarios. This emphasizes the importance of integrating user-centric metrics in the reward function to ensure that maximizing rewards translates into tangible QoE improvements for end-users.
    \item None of the models perform well in \textbf{all} network scenarios, which raises the question of whether it is possible to train a model that performs well across \textbf{all} network conditions. For instance, Schaferct ranks $\#1$ in high bandwidth, burst loss, and fluctuating burst loss scenarios, but fails to be in the top 3 in low bandwidth and fluctuating bandwidth scenarios, where the baseline and the Paramecium models shine, respectively. This can be attributed to the data drift between lab and real-world testbed datasets: low bandwidth and fluctuating bandwidth scenarios are not well represented in the real-world testbed dataset, which has been used to train the Schaferct model. It is also interesting to mention that the Paramecium model ranks among the top $2$ models in low bandwidth, high bandwidth, and fluctuating bandwidth scenarios, but fails to perform competitively in burst loss and fluctuating burst loss. This is expected because Paramecium estimates the bandwidth based on packet delay and rate signals, but ignores the packet loss signal.
\end{enumerate}


In table \ref{tab:firststage}, the average audio and video quality score attained by each model on all network scenarios is reported along with a $95\%$ confidence interval (CI). Schaferct comes in the first place, followed by Paramecium. In the third place, we observe no statistical difference between the SJTU Medialab and the Fast and furious models. Hence, \textbf{Schaferct}, \textbf{Paramecium}, \textbf{SJTU Medialab}, and the \textbf{Fast and furious} models \textbf{were all advanced} to the final evaluation stage on the testbed.

\begin{table}[H]
  \caption{First evaluation stage rankings. }
  \label{tab:firststage}
  \begin{tabular}{|l|l||c|c|c|c|c|c|}
    \hline
    \textbf{Rank} & \textbf{Model} & \textbf{Score} $(\mathbb{S})$ & \textbf{$95\%$ CI} \\
    \hline
    \hline
\textbf{1}	&	\textbf{Schaferct}	&	7.63	&	[7.61, 7.64]	\\  \hline
\textbf{2}	&	\textbf{Paramecium}	&	7.53	&	[7.51, 7.55]	\\  \hline
\textbf{3 (tie)}	&	\textbf{SJTU Medialab}	&	7.51	&	[7.48, 7.55]	\\  \hline
\textbf{3 (tie)}	&	\textbf{Fast and furious}	&	7.51	&	[7.49, 7.53]	\\  \hline
5	&	CUC Echoes	&	7.33	&	[7.28, 7.37]	\\  \hline
6	&	TEN TMS	&	7.32	&	[7.26, 7.38]	\\  \hline
7	&	MediaLab	&	6.50	&	[6.40, 6.60]	\\ 
\hline
\end{tabular}
\end{table}

\subsection{Second Evaluation Stage}

\begin{figure*}[h]
  \centering
  \label{fig:technical}
  \includegraphics[width=\linewidth]{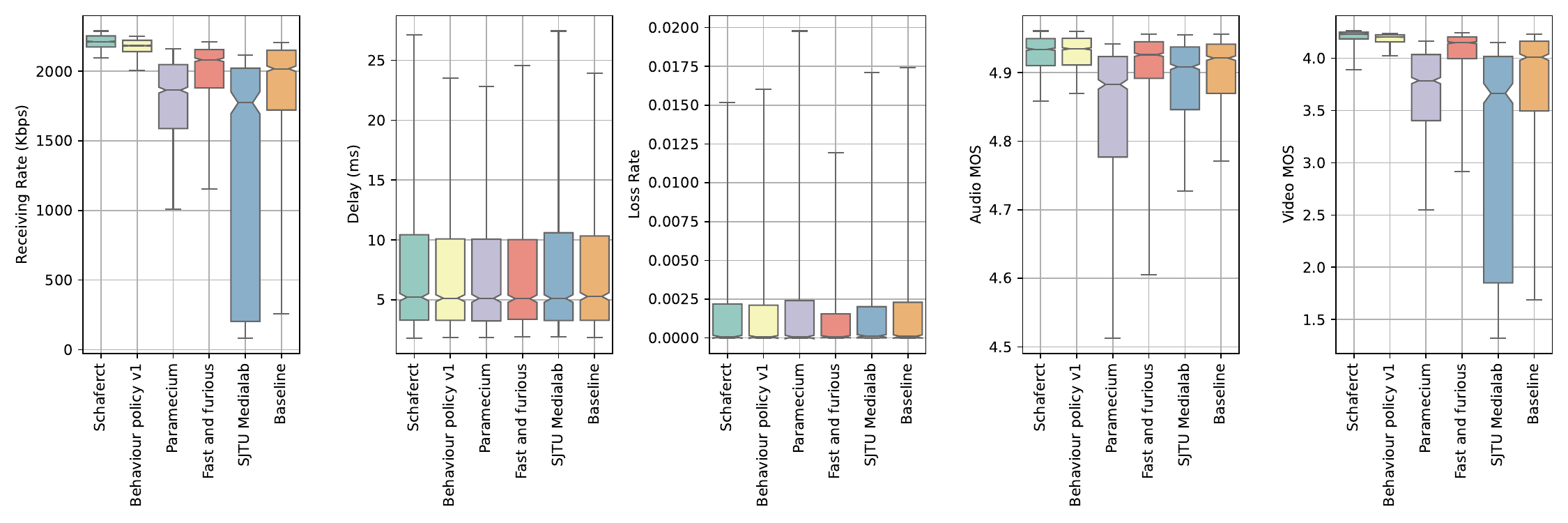}
  \caption{Performance results of top models on the testbed. 
  Each box is based on $1200$ data points. Lower and upper whiskers represent the $10$th and $90$th percentiles, respectively. Model set includes the baseline policy as well as the top behaviour policy in the released datasets (v1). The winning model, Schaferct, demonstrates comparable performance to the best behavior policy (v1) in the released datasets across all metrics.}
\end{figure*}

Based on call data from the final evaluation stage, Figure 1 presents a boxplot of key technical metrics. These metrics include packet receiving rate (Kbps), packet delay (ms), packet loss rate, as well as objective audio and video MOS scores. In addition to assessing the top $4$ models identified in the first evaluation stage, we have also evaluated the baseline model and the top-performing behavior policy within the released datasets (v1). Each box in the boxplot is based on $1200$ data points ($600$ calls $\times$ 2 legs/call). Lower and upper whiskers represent the $10$th and $90$th percentiles, respectively. \textbf{It can be observed that Schaferct demonstrates comparable performance to the best behavior policy (v1) in the released datasets across all metrics, outperforming other models. This provides evidence that offline RL is well-suited for training competitive bandwidth estimators in RTC, by leveraging real-world data and using objective audio/video MOS scores as a reward}. Moreover, it is clear that the receiving rate serves as a reliable predictor for video MOS: models exhibiting a higher receiving rate achieve a correspondingly higher video MOS, while those with lower rates experience the opposite trend. Conversely, the majority of models excel in terms of audio MOS, achieving a median score above 4.9, with the exception of the Paramecium model, whose policy results in a higher packet loss than other models leading to a poorer audio quality.

Last but not least, the rankings of the final evaluation stage are shown in Table \ref{tab:finalstage}. Securing the top position in the grand challenge is the \textbf{Schaferct} model, emerging as the undisputable winner, followed by the \textbf{Fast and Furious} model, which claims the noteworthy status of the runner-up. 

\begin{table}[H]
  \caption{Final evaluation stage rankings. All score deltas are highly statistically significant with p-value < $0.0001$.}
  \label{tab:finalstage}
  \begin{tabular}{|l|l||c|c|c|c|}
    \hline
    \textbf{Rank} & \textbf{Model} & \textbf{Score} $(\mathbb{S})$ & \textbf{95\% CI} \\
    \hline
    \hline
\textbf{1}	&	\textbf{Schaferct}	&	\colorbox{green}{8.93}	&	[8.88, 8.97]	 \\ \hline
\textbf{2}	&	\textbf{Fast and furious}	&	\colorbox{yellow}{8.70}	&	[8.65, 8.76]	 \\ \hline
3	&	Paramecium	&	8.34	&	[8.28, 8.39]  \\\hline
4	&	SJTU Medialab	&	7.89	&	[7.82, 7.96]  \\
\hline
\end{tabular}
\end{table}

\section{Conclusion}

This grand challenge has demonstrated the potential of offline reinforcement learning (RL) to enhance the quality of experience (QoE) for users in real-time communications (RTC). By leveraging a diverse dataset derived from Microsoft Teams calls and incorporating objective audio/video quality scores which have high correlation with subjective experience scores as rewards, the challenge has paved the way for the development of more user-centric bandwidth estimation models. The winning model, which was rigorously tested over a geographically distributed testbed, showcases the effectiveness of the proposed approach for designing new bandwidth estimators. The insights gained from this challenge will undoubtedly inform future research and development efforts aimed at optimizing QoE for users across diverse network conditions. As we continue to explore the capabilities of offline RL, it is imperative that we maintain a focus on user-centric metrics, ensuring that technological advancements translate into tangible benefits for end-users.

\newpage

\bibliographystyle{ACM-Reference-Format}
\bibliography{references}

\end{document}